# Gd-Based Solvated Shells for Defect Passivation of $CsPbBr_3$ Nanoplatelets Enabling Efficient Color-Saturated Blue Electroluminescence

Haoran Wang[1,3,4], Jingyu Qian[6], Jiayun Sun[3,4,5], Tong Su[1], Shiming Lei[7], Xiaoyu Zhang[6], Wallace C. H. Choy[5], Xiao Wei Sun[3,4], Kai Wang[3,4]*, Weiwei Zhao[1,2]*

[1] Sauvage Laboratory for Smart Materials, The School of Materials Science and Engineering, and Shenzhen Key Laboratory of Flexible Printed Electronics Technology, Harbin Institute of Technology, Shenzhen 518055, China

[2] State Key Laboratory of Advanced Welding & Joining, Harbin Institute of Technology, Harbin 150001, China

[3] Institute of Nanoscience and Applications, and Department of Electrical and Electronic Engineering, Southern University of Science and Technology, Shenzhen 518055, China

[4] Key Laboratory of Energy Conversion and Storage Technologies, Southern University of Science and Technology, Ministry of Education, Shenzhen 518055, China

[5] Department of Electrical and Electronic Engineering, The University of Hong Kong, Pokfulam Road, Hong Kong, China

[6] College of Materials Science and Engineering, Jilin University, Changchun 130012, China

[7] Department of Physics, The Hong Kong University of Science and Technology, Clear Water Bay, Kowloon, Hong Kong, China

*E-mail: wzhao@hit.edu.cn and wangk@sustech.edu.cn

## ABSTRACT:

Reduced-dimensional $CsPbBr_3$ nanoplatelets (NPLs) are promising candidates for color-saturated blue emitters, yet their electroluminescence performance is hampered by non-radiative recombination, which is associated with bromine vacancies. Here, we show that a post-synthetic treatment of $CsPbBr_3$ NPLs with $GdBr_3$-dimethylformamide (DMF) can effectively eliminate defects while preserving the color. According to a combined experimental and theoretical study, $Gd^{3+}$ ions are less reactive with NPLs as a result of compact interaction between them and DMF, and this stable $Gd^{3+}$-DMF solvation structure makes $Br^-$ ions more available and allows them to move more freely. Consequently, defects are rapidly passivated and photoluminescence quantum yield increases dramatically (from 35 to ~100%), while the surface ligand density and emission color remain unchanged. The result is a remarkable electroluminescence efficiency of 2.4% (at 464 nm), one of the highest in pure blue perovskite NPL light-emitting diodes. It is noteworthy that the conductive NPL film shows a high photoluminescence quantum yield of 80%, demonstrating NPLs' significant electroluminescence potential with further device structure design.

Metal halide perovskites exhibit high color purity, facile tunable color, and excellent carrier mobility, making them promising candidates for next-generation light-emitting diodes (LEDs)[1-7]. Significant progress has been made in improving the performance of perovskite LEDs (PeLEDs) in the past few years. At present, the external quantum efficiencies (EQEs) reported for green-, red-, and near-infrared-emitting PeLEDs have reached over 20%[8-13]. However, their pure-blue counterparts (with emission around 465 nm) are still far behind[14-16], hindering their practical applications in full-color displays and solid-state lighting.

Generally, there are two approaches to realizing blue-emitting perovskites. One is to incorporate Cl into Br-based perovskite to enlarge its bandgap[1-4]. However, there are two critical limitations impeding the practical implementation of Cl:Br mixed perovskites[17,18]: (i) Cl vacancies are easily formed, resulting in poor photoluminescence quantum yield (PL QY); (ii) inevitable phase segregation due to halide migration under electrical pressure induces serious color shift or multi-peaks. The other one is to utilize the quantum confinement effects in pure-Br perovskites [i.e., quasi-two-dimensional (quasi-2D) structure, ultrasmall quantum dots, and 2D nanoplatelets (NPLs)][14,16,19-24]. In particular, NPLs with narrow band-edge emission and large exciton binding energies have shown promising features for spectral-stable blue-emitting active layers[20-26]. However, it has been noted that LEDs made with blue NPLs give very limited EL performance[23,27-32]. The reason could be described as follows. NPLs have a high surface-to-volume ratio and so numerous surface defects[25,30], and these can facilitate severe non-radiative recombination of charge carriers, resulting in low PL QYs and ultimately low device performance. Surface defects are generally believed to be related to Br vacancies[30]. Surface treatments, by using polymers and organic ligands, have been applied to passivate these vacancies in NPLs and thus increase PL QYs, but at the expense of their electrical conductivity[27,28,33], making them not suitable for constructing high-performance LEDs. Consequently, a feasible strategy is vital for improving luminescence efficiency while retaining decent electrical properties.

Post-synthetic passivation with metal halide salts can render NPL films with high carrier mobilities. Meanwhile, they also improved the PL QY by healing of the halide vacancy defects, as demonstrated in perovskite nanocrystals[34-38]. Such a strategy, however, is not feasible for NPLs since small metal ions diffuse and are easily incorporated into the crystal lattice, reducing the growth energy barrier, resulting in multiple emissions or even green emission from large-sized nanocrystals grown from NPLs[39,40]. Furthermore, due to the strong affinity of metal ions for NPL surfaces, they were able to enhance the dynamic detaching of the native capping ligands. Insufficient ligands on NPLs allow these particles to aggregate together and grow into large dimensions[24], which aggravates the above disadvantages.

In this work, we show how to tackle this issue by combining rational metal bromide selection with solvation engineering. We demonstrate a post-synthetic treatment with $GdBr_3$-dimethylformamide (DMF) that improves the PL QY of $CsPbBr_3$ NPLs to near unity, while simultaneously maintaining the desired emission color. Several surface characterizations as well as density functional theory (DFT) calculations were performed to study the reaction mechanism of $GdBr_3$-DMF post-treatment. The results demonstrate that the compact interaction between the $Gd^{3+}$ ions and DMF molecules causes them to move within the first solvation shell instead of moving freely. In other words, the $Gd^{3+}$ ion acts as an inert bystander during the post-synthetic process, which has a negligible effect on the NPL's surface composition and its emission color. Meanwhile, this stable $Gd^{3+}$-DMF solvation structure imparts adequate freedom of movement to Br⁻ ions, which leads to rapid defect passivation and a dramatic increase in PL QYs. As a result, we are able to realize a remarkable electroluminescence efficiency of 2.4% (at 464 nm), which represents one of the highest efficiencies in pure blue PeLEDs based on NPLs reported to date.

For this investigation, blue-emitting $CsPbBr_3$ NPLs were synthesized following our previous report with minor modifications (see the "Experimental Section")[25]. The isolated as-synthesized $CsPbBr_3$ NPLs were first examined by X-ray diffraction (XRD), indicating they are orthorhombic structures (**Figure S1a**). A transmission electron microscopy (TEM) image of NPLs in the inset of **Figure S1a** exhibits the typical platelet-shaped nanostructure with a mean lateral size of 12.8 nm (**Figure S1b**). The optical characterizations reveal an absorption peak at 449 nm and a blue emission at 460 nm with a PL QY of 35.1% (**Figure S1c**). Such a low PL QY value has commonly been observed and is related to a high density of Br vacancy defects in NPLs[30]. Inorganic metal bromides can eliminate surface defects by providing a Br-rich environment, and simultaneously increase the electrical properties of NPLs. We screened three metal bromides (NaBr, $ZnBr_2$, and $GdBr_3$) that possess different Br contents and binding affinities with $CsPbBr_3$ NPLs for post-synthetic passivation. These metal bromides were first dissolved in dimethylformamide (DMF, 0.1 M), and 2 μL of each metal bromide solution was then added to 1 mL of ethyl acetate. After that 0.5 mL of pristine NPLs was added to the above solution. The mixture was vortexed for 15 s and then centrifuged (see the Experimental Section for details).

A rapid (<3 s) change of both solution color and fluorescence is observed upon the addition (**Figure S2**). The reaction of the NPLs with NaBr-DMF and ZnBr$_2$-DMF leads to pronounced red-shifts in the PL spectra, with an emission maximum of around 510 nm for both samples (**Figure 1a**). We attribute this shift to the regrowth of CsPbBr$_3$ NPLs and the subsequent loss of quantum confinement. It is well-known that the commonly used acid/amine ligands are intrinsically mobile and easily detached from perovskite surfaces[41]. Furthermore, thanks to the strong affinity of metal ions with the NPL, they could enhance the dynamic detaching of ligands (as will be discussed in detail later) and thus promote regrowth into large-size particles via bare-surface contact. The morphologic changes were confirmed by TEM images (**Figure 1b**, **1c**), demonstrating that the NPLs evolve into thicker multilayer structures (with NaBr-DMF treatment) and fused nanocubes (with ZnBr$_2$-DMF treatment).

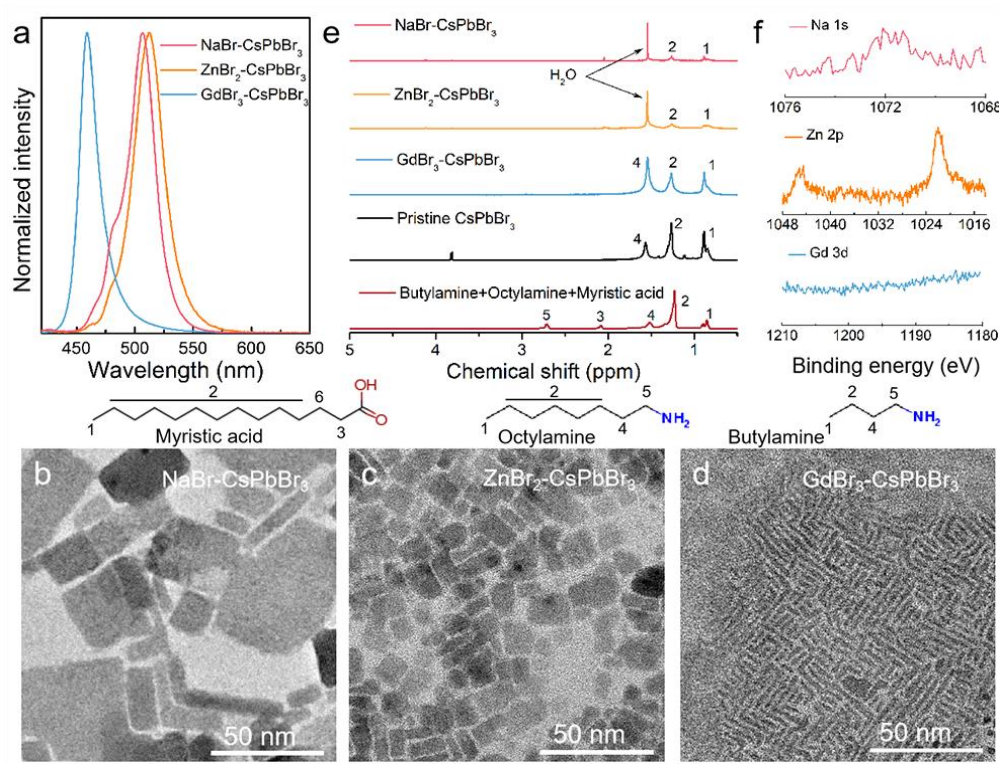

**Figure 1**. (a) PL spectra of CsPbBr$_3$ NPLs after reaction with NaBr-DMF, ZnBr$_2$-DMF, and GdBr$_3$-DMF (reaction time: 15 s). TEM images of CsPbBr$_3$ NPLs after reaction with (b)

NaBr-DMF, (c) ZnBr$_2$-DMF, and (d) GdBr$_3$-DMF. The average lateral size of NaBr-, ZnBr$_2$-, and GdBr$_3$-CsPbBr$_3$ is 27.6, 15.9, and 13.1 nm, respectively (**Figure S3**). (e) H$^1$NMR of butylamine+octylamine+myristic acid (the ligands employed for NPL synthesis), pristine CsPbBr$_3$, and metal bromides-treated CsPbBr$_3$, in CDCl$_3$. H$^1$NMR spectra of pristine CsPbBr$_3$ NPLs reveal the absence of myristic acid on their surface, with only protonated amine species being detected (**Figure S7**). The absence could be due to the fact that during the purification process, the surface-bound myristic acid was protonated by the acetic acid molecules and thus detached from the NPLs surface[42]. High-resolution (f) XPS spectra of NaBr-CsPbBr$_3$ for Na 1s, ZnBr$_2$-CsPbBr$_3$ for Zn 2p, and GdBr$_3$-CsPbBr$_3$ for Gd 3d.

Notably, we find that the morphologies of GdBr$_3$-NPLs (**Figure 1d**), as well as the PL shape, are similar to those of the pristine NPLs; this implies that the GdBr$_3$-DMF treatment does not greatly influence the ligands bound on the NPL surface. At the same time, the absolute PL QY of GdBr$_3$-NPLs was measured near 100%, which is a 2.8-fold increase as compared to that of the pristine NPL. This significant improvement indicates that surface defects are effectively reduced following GdBr$_3$-DMF treatment. Similarly, the PL decay (**Figure S4** and **Table S1**) of the GdBr$_3$-NPLs (45.9 ns) is slower than that of the pristine NPLs (18.7 ns), further confirming the efficient passivation of defects. Importantly, the GdBr$_3$-NPLs exhibit a high blue-emitting PL QY of 80% in solid-state films (**Figure S5**), which is one of the highest values in blue perovskite NPL films (**Table S2**). This is contrary to the general expectation that spin-coated films from purified NPLs have poor color purity and low

PL QYs (because surface ligands are lost during purification). We attribute this to the synthetic method employed[25], as well as the Br-rich environment provided by $GdBr_3$, which can efficiently preserve the original surface properties of the treated NPLs during purification. **Figure S6a** shows PL QYs and PL peak positions of NPLs treated with different concentrations of $GdBr_3$ (from 0.03 to 0.1 M). PL emission peaks are barely changed at ~ 460 nm, but it clearly shows that the PL QY increases as a function of the amount of $GdBr_3$ used. In contrast, both NaBr- and $ZnBr_2$-$CsPbBr_3$ show a significant red-shift in PL peaks and a limited improvement in PL QYs (or even worsening) at any given concentration (**Figure S6b**, **S6c**). Combined with the above, it is reasonable to speculate that $GdBr_3$-DMF is able to provide more available $Br^-$ ions to repair the surface, while the $Gd^{3+}$ ion acts as an inert bystander during the post-synthetic process and thus has no effect on PL.

To verify this hypothesis, we comparatively studied the surface composition and chemical states of NaBr-, $ZnBr_2$-, and $GdBr_3$-$CsPbBr_3$ by $^1H$ nuclear magnetic resonance spectroscopy (NMR) and X-ray photoelectron spectroscopy (XPS). From $^1H$ NMR, the surface composition of the $CsPbBr_3$ NPLs was found to remain stable after the reaction with $GdBr_3$-DMF. In contrast, a strong and narrow peak at ~1.54 ppm derived from water molecules appears both in NaBr-$CsPbBr_3$ and $ZnBr_2$-$CsPbBr_3$, indicating that a lot of water molecules are adsorbed on the $CsPbBr_3$ surface, which was confirmed again by FTIR spectra (**Figure S8**). It is proposed that hydration reactions can occur easily on account of the removal of surface ligands[43]. Indeed, quantitative XPS analysis shows a significant reduction in capping ligand density after NaBr-DMF and ZnBr$_2$-DMF treatment; the N/Pb ratios of pristine, NaBr-, and ZnBr$_2$-CsPbBr$_3$ were calculated to be 1.32, 0.9, and 0.63, respectively. While this ratio was 1.24 for GdBr$_3$-CsPbBr$_3$, which is quite close to pristine CsPbBr$_3$, reconfirming that GdBr$_3$-DMF treatment does not induce a significant change in the surface ligand composition. On the other hand, the Br/Pb ratio of GdBr$_3$-CsPbBr$_3$ (4.02) was increased as compared to that of pristine CsPbBr$_3$ (3.51), demonstrating that Br species in GdBr$_3$ are incorporated on the NPL surface. XPS results also indicate the presence of Na and Zn elements respectively for NaBr-CsPbBr$_3$ and ZnBr$_2$-CsPbBr$_3$ (**Figure 1f**). If there was doping, the doping effects would induce the red-shift in diffraction peaks and blue-shift in PL emission peaks since the ionic radius of Na$^+$ (Zn$^{2+}$) is smaller than that of Cs$^+$ (Pb$^{2+}$) (Na$^+$: 1.02 Å; Cs$^+$:1.67 Å; Zn$^{2+}$: 0.74 Å; Pb$^{2+}$: 1.19 Å). However, we did not observe these phenomena (XRD, **Figure S9**), and therefore concluded that Na$^+$ (Zn$^{2+}$) ions are mainly bonded on the CsPbBr$_3$ surface after partially stripping the native capping ligands. For GdBr$_3$-CsPbBr$_3$, no Gd signal was detected, which suggests that Gd$^{3+}$ ions are absent from the CsPbBr$_3$ surface. Inductively coupled plasma mass spectrometry (ICP-MS) suggests no sign of the doping of Gd$^{3+}$ ions into the crystal lattice (below the detection limit). Although the Gd$^{3+}$ ion can be incorporated into the crystal lattice by the hot-injection method[44], the probability of introducing it into our samples is low under the reaction conditions (short reaction time (15 s) at room temperature (approx. 25 °C)).

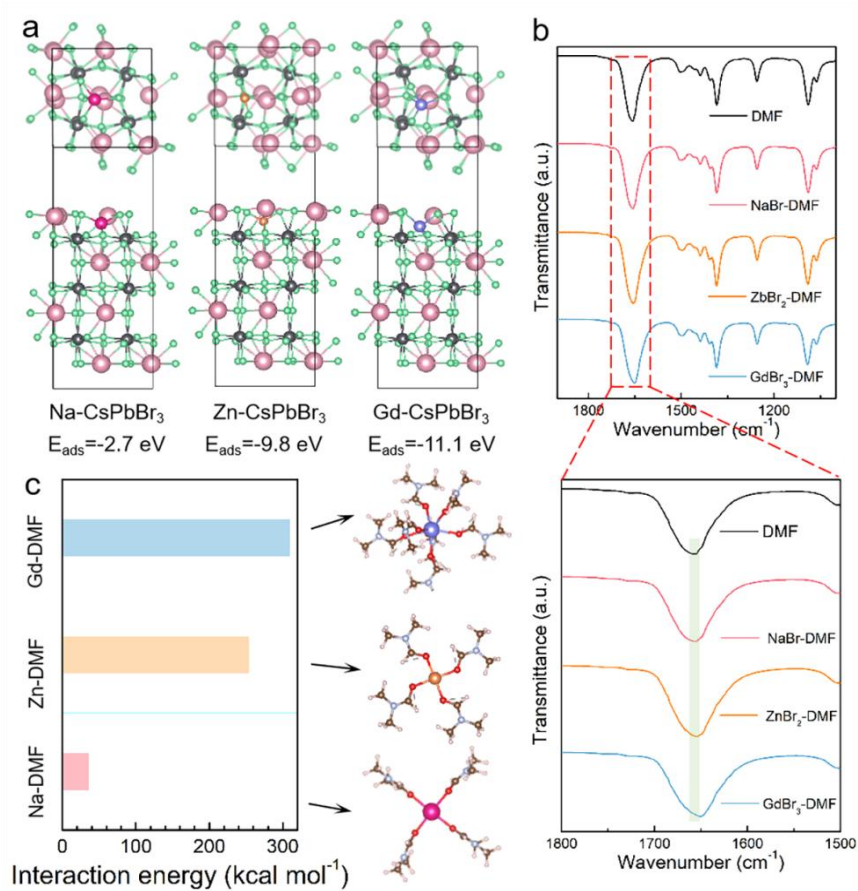

**Figure 2**. (a) Adsorption energy of $Na^+$-, $Zn^{2+}$-, and $Gd^{3+}$-capped surfaces (top and side views). (b) Fourier transform infrared (FTIR) of Na-, Zn- and Gd-DMF (0.1 M). (c) The interaction energy in the first solvation shell of $Na^+$-DMF, $Zn^{2+}$-DMF, and $Gd^{3+}$-DMF, and the corresponding schematic of the solvation shell.

Density functional theory (DFT) calculations are performed to investigate the adsorption behaviors of $Na^+$, $Zn^{2+}$ and $Gd^{3+}$ ions on the $CsPbBr_3$ surfaces. As shown in **Figure 2a**, the $Gd^{3+}$-capped surface has larger adsorption energy ($E_{ads}$, -11.1 eV) than that of $Na^+$- (-2.7 eV) and $Zn^{2+}$-capped surface (-9.8 eV). The larger $E_{ads}$ indicate a stronger binding affinity of $Gd^{3+}$ ions to the $CsPbBr_3$ NPLs, which is more effective in promoting surface ligand detachment, in obvious conflict with our above observations. A reason for this is the oversimplification of the calculations. Our approach ignores the effects of solvents and assumes that metal ions have adequate freedom of movement. Among the solvents used, DMF possesses the highest Gutmann donor number (26.6) and dielectric constant (36.7)[45]. These values suggest that DMF can serve as a suitable solvent that coordinates with metal ions. The presence of DMF-metal ion complexes would restrict the movement of metal ions and thus the adsorption process. Considering the fact mentioned above, we turned our attention to the reaction between these metal bromides and DMF, and the resulting solvation structure of metal ions.

The interaction between DMF with metal ions was demonstrated by FTIR spectroscopy (**Figure 2b**). For the bare DMF, the peak located at 1657 cm$^{-1}$ represents the C=O bond stretching vibration. The C=O peak downshifts to 1655 cm$^{-1}$, 1654 cm$^{-1}$, and 1651 cm$^{-1}$ after the addition of NaBr, ZnBr$_2$, and GdBr$_3$, respectively. The such shift indicates the formation of DMF-metal ion complexes through the Na$^+$ (Zn$^{2+}$or Gd$^{3+}$)-O bond. Since the magnitude of the shift in C=O stretching vibration frequency is correlated with the strength of DMF-metal ions interaction, the strength of the interaction between the DMF and metal ions would be the highest for Gd$^{3+}$ ions followed by Zn$^{2+}$ ions and Na$^+$ ions.

Further, the interaction energy of the first solvation shell was compared by DFT calculations. As shown in **Figure 2c**, the first solvation shell of Gd$^{3+}$ ions involves seven DMF molecules, with an interaction energy of 308.31 kcal mol$^{-1}$, while that of Na$^+$ ions and Zn$^{2+}$ ions only consists of four DMF molecules, with an interaction energy of 34.37 kcal mol$^{-1}$ for Na-DMF and 252.74 kcal mol$^{-1}$ for Zn-DMF. Due to the tight interaction between DMF molecules and Gd$^{3+}$ ions, they move within the first solvation shell instead of having free movement. In contrast, both Na$^+$ and Zn$^{2+}$ ions are relatively easy to escape the first solvation shell due to the looser solvation structure and lower interaction energies. Consequently, there would be more free metal ions that facilitate surface ligand detachment. We note that although the interaction energy of Zn-DMF is much higher than that of Na-DMF, the red shift of the PL peak is more pronounced in Zn-CsPbBr$_3$. In addition to surface ligand detachment, other mechanisms may contribute to PL red-shifting in Zn-sample. For example, a previous report showed that the incorporation of Zn$^{2+}$ ions can reduce the growth energy barrier in growing green CsPbBr$_3$ nanocrystals from blue CsPbBr$_3$ NPLs[39]. It enables the layer-by-layer growth reaction on the basis of pristine platelets, accompanied by a significant red-shift in PL spectra. In our study, very similar observations have been made after the introduction of ZnBr$_2$, but no obvious shift of diffraction peak was observed in the corresponding XRD patterns; this could be understood considering the low doping concentration of Zn$^{2+}$ ions.

 Based on these results, we propose a mechanism for the post-synthetic treatment of CsPbBr$_3$ NPLs using Gd-DMF and Na (Zn)-DMF, as shown in **Figure 3**. According to the Stern theory of the electric double layer, there is a less-compact diffusion layer (the secondary solvation shell) outside the first solvation shell[46]; it may be composed of the attracted DMF molecules (via van der Waals force), free DMF molecules, the attracted Br ions (via van der Waals force), free Br$^-$ ions and the ions pair. The compact interaction between DMF molecules and $Gd^{3+}$ ions could decrease the number of free DMF molecules, in turn reducing the detrimental effects of DMF on NPL surface properties. Also, this stable Gd-DMF solvation structure makes $Br^-$ ions less likely to contact $Gd^{3+}$ ions to form ion pairs (larger size and mass being less active). Therefore, $Br^-$ ions have adequate freedom of movement and high reactivity, which enables rapid defect passivation and a dramatic increase in PL QY.

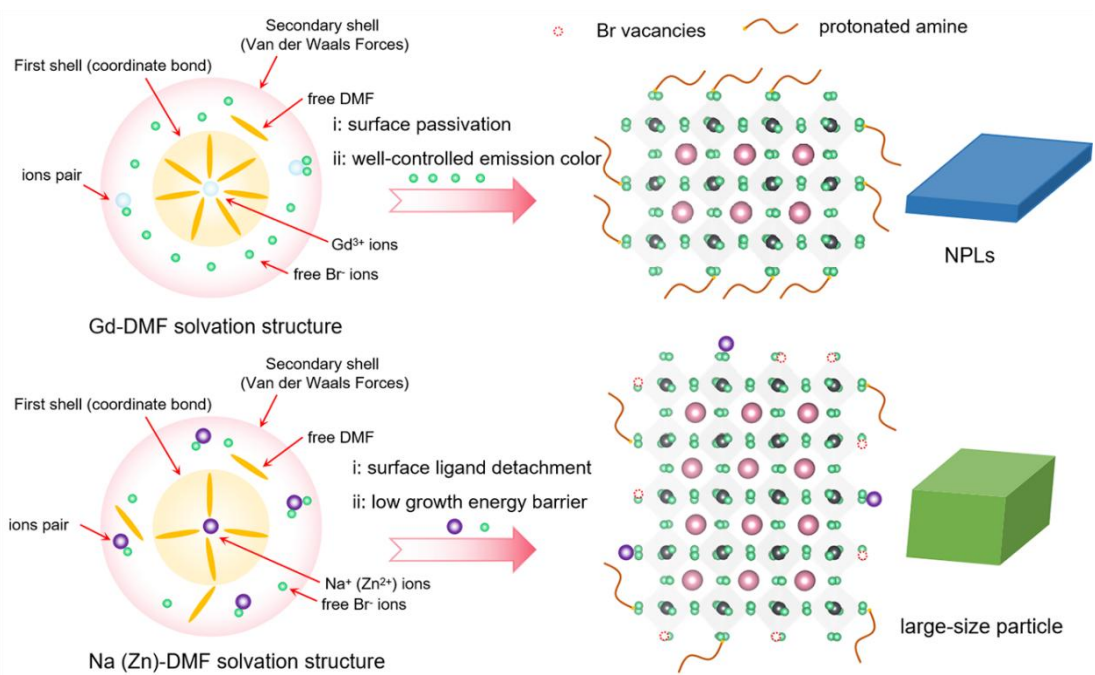

**Figure 3**. Schematic of post-synthetic treatment of $CsPbBr_3$ NPLs using Gd-DMF and Na (Zn)-DMF.

To assess the suitability of $GdBr_3$-treated $CsPbBr_3$ NPLs as light emitters for LEDs, a device with a multilayer architecture of ITO/PEDOT:PSS/PTAA/PEABr/$CsPbBr_3$ NPLs/TPBI/LiF/Al was fabricated (**Figure 4a**), for which the corresponding energy level diagram is shown in **Figure 4b**. Devices using pristine $CsPbBr_3$ NPLs as emitters were also fabricated for comparison.

The EL spectra for the two types of devices show a center emission wavelength at ~ 464 nm (**Figure 4c**). The GdBr$_3$-treated device shows a reduction in EL FWHM, namely 16 nm as compared to 17 nm for the pristine-based device, which is attributed to the reduction of the band-edge shallow-level defects. We also note that a relatively weak EL emission shoulder at ~ 510 nm was recorded for the device based on pristine CsPbBr$_3$ NPLs. We believe that these impurity emissions originate from the local aggregation of CsPbBr$_3$ NPLs under the action of an electric field. On the contrary, GdBr$_3$-treated NPL devices present a single emission wavelength without any shoulder peak. These results prove that GdBr$_3$ treatment is able to provide NPLs with robust stability, thereby effectively preventing electric-field-induced phase transformation. Thus, the GdBr$_3$-treated NPLs exhibit higher colour purity over the pristine NPLs, with CIE coordinates of (0.14, 0.07), indicating very pure blue emission, which is desirable for display applications.

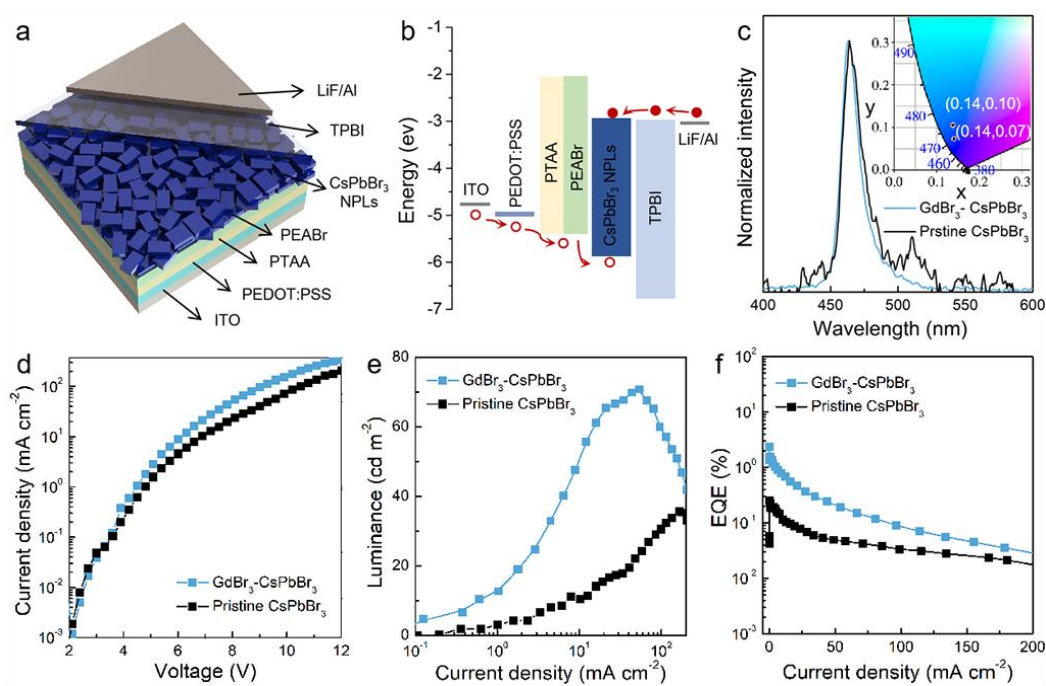

**Figure 4**. (a) Device structure of the PeLED. (b) Schematic flat-band energy diagram of the

PeLED; the energy levels for PEDOT:PSS, PTAA, CsPbBr$_3$ NPLs, and TPBI are taken from reference[25]. (c) Normalized EL spectra and the corresponding Commission Internationale de l'Echlaiage (CIE) color coordinates (inset) of the LED devices based on pristine and GdBr$_3$-CsPbBr$_3$ NPLs at an applied voltage of 8.1 V. (d) Current density-voltage (*J-V*), (e) luminance-*J* (*L-J*) and (f) external quantum efficiency-*J* (*EQE-J*) curves of the devices based on pristine and GdBr$_3$-treated CsPbBr$_3$ NPLs.

The EL characteristics for the resulting PeLEDs are plotted in **Figure 4d-f**. From the *J–V* curves, it is found that the current injection is more efficient for the device based on GdBr$_3$-treated NPL. The turn-on voltage of the GdBr$_3$-NPL LED is 3.3 V, lower than that of devices that rely on pristine NPLs (4.2 V). The maximum luminance (71 cd m$^{-2}$) of the LED fabricated on the GdBr$_3$-treated NPL was nearly two times higher than that of the pristine NPL device (36 cd m$^{-2}$). The device based on pristine NPL shows poor luminous efficiency (max. EQE: 0.25%). In contrast, the device using GdBr$_3$-treated NPL shows a peak EQE of 2.4%, which represents one of the highest efficiencies in pure blue PeLEDs based on NPLs reported to date (**Table S3**). **Figure S10** provides a histogram of the maximum EQE values for 10 devices built on GdBr$_3$-treated NPLs. The average maximum EQEs are 2.01%, with a relative standard deviation of 8.74%, suggesting excellent performance reproducibility. The results demonstrate a very promising method to fabricate NPL-based PeLEDs featuring high performance and superior spectral stability.

In summary, we have developed an effective post-synthetic strategy toward highly luminescent and stable $CsPbBr_3$ NPLs for efficient blue LED devices. We have shown that post-synthetic treatment of $CsPbBr_3$ NPLs with $GdBr_3$-DMF is able to very effectively passivate the defects generated by bromine vacancies, while simultaneously maintaining the desired emission color. We have found that the $GdBr_3$-DMF treatment is unique compared to surface treatments using NaBr-DMF or $ZnBr_2$-DMF. Our results suggest that $Gd^{3+}$ ions exhibit a stronger interaction and tighter connection with DMF compared to $Na^+$ and $Zn^{2+}$ ions; the less-reactive $Gd^{3+}$-DMF has a negligible effect on the NPL's surface composition and its emission color. Meanwhile, this stable $Gd^{3+}$-DMF solvation imparts adequate freedom of movement to $Br^-$ ions, which leads to rapid defects passivation and a dramatic increase in luminous efficiency (PL QY from 20 to ~100%). As a result, we are able to realize a remarkable electroluminescence efficiency of 2.4% (at 464 nm), which represents one of the highest efficiencies in pure blue PeLEDs based on NPLs reported to date. Our findings provide a new way to tailor the optical properties of perovskite NPLs by rational metal bromide selection and solvation engineering.

## ASSOCIATED CONTENT

**Supporting Information.** The Supporting Information including experimental methods, characterizations, supplementary figures, and supplementary tables are available free of charge on the ACS Publications website.

## AUTHOR INFORMATION

### Corresponding Author

wzhao@hit.edu.cn

wangk@sustech.edu.cn

**Author Contributions**

H.W. performed the experiment and wrote the manuscript. J.Q. carried out density functional theory calculations. H.W., J.Q., J.S., T.S., S.L., X.Z., W.C., X.W.S., K.W. and W.Z. analyzed and discussed the experimental results. All authors contributed to the manuscript.

**Notes**

The authors declare no competing financial interests.

# ACKNOWLEDGMENTS

The authors would like to thank the financial support from the National Key Research and Development Program of China (2021YFB3602703, 2022YFB3606504, and 2022YFB3602903), Natural Science Foundation of China (52073075, 62122034, 61405089, and 62005115), Shenzhen Science and Technology Program (KQTD20170809110344233 and JCYJ20210324104413036), Key-Area Research and Development Program of Guangdong Province (2019B010925001 and 2019B010924001), Shenzhen Key Laboratory for Advanced Quantum Dot Displays and Lighting (ZDSYS201707281632549), Shenzhen Innovation Project (JCYJ20220818100411025), and General Research Fund (Grant Nos. 17201819 and 17211220) and Collaboration Research Fund (C7035-20G) from Hong Kong Special Administrative Region, China.

**TOC Image**

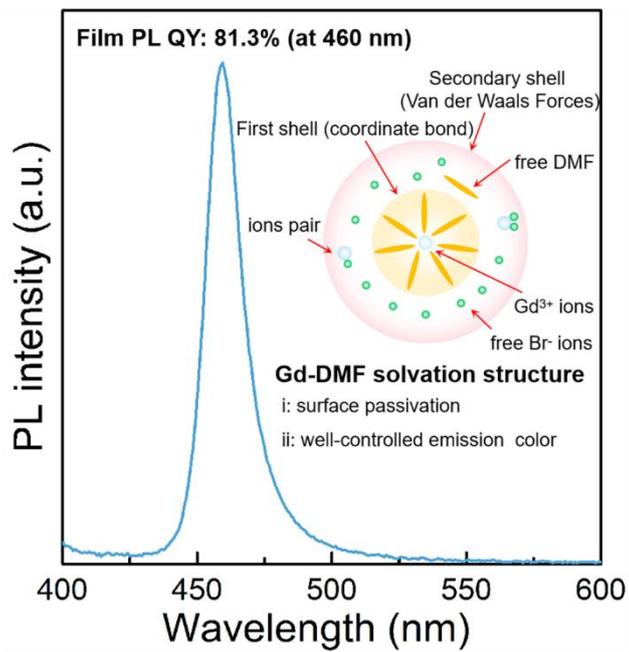